\documentstyle[sprocl,amsmath]{article}

\input{psfig}

\newcommand{\etal}{\emph{et al.}}
\begin{document}

\title{CLUSTERS IN VARIOUS COSMOLOGICAL MODELS:
ABUNDANCE AND EVOLUTION\footnote{To be published in
``Large-Scale Structures: Tracks and Traces. Proceedings
of the 12th Potsdam Cosmology Workshop'', eds. V. M\"uller,
S. Gottl\"ober, J.P. M\"ucket, J. Wambsganss.}}

\author{ M.A.K. GROSS, R.S. SOMERVILLE, J.R. PRIMACK }

\address{ Physics Department, University of California, Santa Cruz, CA 95064,
          USA }

\author{ S. BORGANI }

\address{ INFN Sezione di Perugia, c/o Dipartimento di Fisica dell'Universit\`a,
          via A. Pascoli, I-06100 Perugia, Italy }

\author{ M. GIRARDI }

\address{ SISSA, via Beirut 4, 34013 Trieste, Italy }


\maketitle\abstracts{
The combination of measurements of the local abundance of rich clusters of
galaxies and its evolution to higher redshift offers the possibility of a
direct measurement of $\Omega_0$ with little contribution from other
cosmological parameters.  We investigate the significance of recent claims
that this evolution indicates that $\Omega_0$ must be small.  The most
recent cluster velocity dispersion function \cite{girardi:1997a} from
a compilation including the ESO
Northern Abell Cluster Survey (ENACS) results in a significantly higher
normalization for models, corresponding to $\sigma_8\approx 0.6$ for
$\Omega_0=1$, compared to the Eke, Cole, \& Frenk \cite{eke:1996} result of
$\sigma_8=0.52\pm 0.04$.  Using the ENACS data for a $z=0$ calibration results
in strong evolution in the abundance of
clusters, and we find that the velocity dispersion function is consistent with
$\Omega_0=1$.  The results are dependent upon the choice and analysis
of low-redshift and high-redshift data, so at present, the data is not good
enough to determine $\Omega_0$ unambiguously.}

%%%%%%%%%%%%%%%%%%%%%%%%%%%%%%%%%%%%%%%%%%%%%%%%%%%%%%%%%%%%%%
% Article body                                               %
%%%%%%%%%%%%%%%%%%%%%%%%%%%%%%%%%%%%%%%%%%%%%%%%%%%%%%%%%%%%%%

\section{Introduction}
The present-day number density of rich clusters of galaxies and its evolution
to high redshift can in principle provide a sensitive estimate of the value
of $\Omega_0$.  Several recent papers have attempted
measurements of $\Omega_0$ from the evolution in cluster number density, as
estimated from observations of X-ray temperatures,%
\cite{eke:1996,viana:1996,henry:1997} X-ray luminosities,%
\cite{henry:1992,burke:1997} and virial masses.%
\cite{borgani:1997a,carlberg:1997a,bahcall:1997}

The Press-Schechter \cite{press:1974} approximation accurately represents the
number density of
fairly massive clusters in $N$-body simulations.  It has a Gaussian cutoff
whose position is controlled roughly by $\delta_{\rm c}/D\sigma_8$ where $D$
is the linear fluctuation growth factor, so a small error in
either $\delta_{\rm c}$ or $\sigma_8$ can lead to a misestimate of the amount of
evolution.  For this reason, we perform a careful
renormalization of models to $z=0$ cluster data, including calibration of
$\delta_{\rm c}$ against high-resolution simulations.\cite{gross:suites}

\section{How to Use the Press-Schechter Approximation}
The basic strategy is to relate a directly observable quantity to the virial
mass, and then use the standard Press-Schechter formalism to obtain an
abundance for that virial mass.  For velocity dispersions observations, we
assume that clusters are virialized and collapse spherically.

Girardi \etal\cite{girardi:1997a} (hereafter G97) calculated their preliminary
mass function
using virial radii that were estimated roughly assuming a low-$\Omega_0$
cosmology.  Because of that, they need correction for other models.  This
requires knowledge of the mass (or $\sigma_{\rm 1D}$) profile, and we adopt the
Navarro, Frenk, \& White \cite{navarro:1996b} (NFW) $\sigma_{\rm 1D}$ profile
with $c\equiv 7$.

Once one has calculated a virial mass, one gets the abundance using the
standard Press-Schechter formalism.  This requires an assumption about the
critical linear density for collapse ($\delta_{\rm c}$).  We calibrated that
from simulations, where the masses were measured from the simulations using
similar techniques to the real data --- that is, for G97 masses, measuring
the mass within the ``virial'' radius as G97 defines it, and $\delta_{\rm c}$ is
tuned so that the G97-mass Press-Schechter formalism agrees with simulations
at $5.5\times 10^{14}$ $h^{-1}$ $M_\odot$.

We used this formalism to find values of $\sigma_8$ and $n$ that
produce a minimum $\chi^2$ subject to the Bunn \& White \cite{bunn:1997}
\emph{COBE} normalization.  The results are shown in table~\ref{tbl:sig8}.
\begin{table}[htb]
  \caption{Model renormalizations to the G97 mass function and
           four year \emph{COBE} normalization, compared to Eke, Cole, \& Frenk.
           \protect\cite{eke:1996}}
\footnotesize\centering
\begin{tabular}{|l*{8}{r@{.}l}r@{.}c@{$\pm$}c@{.}l@{}l|}
\hline
Model&
\multicolumn{2}{c}{$\Omega_{\rm c}$}&
\multicolumn{2}{c}{$\Omega_{\rm b}$}&
\multicolumn{2}{c}{$\Omega_{\rm h}$}&
\multicolumn{2}{c}{$\Omega_\Lambda$}&
\multicolumn{2}{c}{$h$}&
\multicolumn{2}{c}{$\delta_{\rm c}$}&
\multicolumn{2}{c}{$n$}&
\multicolumn{2}{c}{$\sigma_8$}&
\multicolumn{4}{c}{$\sigma_{\rm 8,ECF}$}&\\
\hline
CHDM-2$\nu$ & 0&731 & 0&069 & 0&2 & 0&0 & 0&6 & 1&357 & 0&912 & 0&589 &
        0&52&0&04&\\
$\Lambda$CDM& 0&331 & 0&069 & 0&0 & 0&6 & 0&6 & 1&400 & 1&000 & 0&805 &
        0&80&0&06&\\
OCDM        & 0&431 & 0&069 & 0&0 & 0&0 & 0&6 & 1&466 & 0&907 & 0&837 &
        0&69&0&05&\\
TCDM        & 0&900 & 0&100 & 0&0 & 0&0 & 0&5 & 1&316 & 0&864 & 0&574 &
        0&52&0&04&\\
\hline
\end{tabular}
  \label{tbl:sig8}
\end{table}

\section{Evolution of Cluster Abundances and Cosmological Parameters}
Now that we have refined the normalization parameters to correspond to $z=0$
velocity dispersion data, we can use the calibrated Press-Schechter formalism
to extrapolate each model's mass function to higher redshift.  Then, we can
compare to high redshift velocity dispersion data from the Canadian Network
for Observational Cosmology \cite{carlberg:1997a} (CNOC), to discriminate
between models.

\begin{figure}[htb]
  \psfig{figure=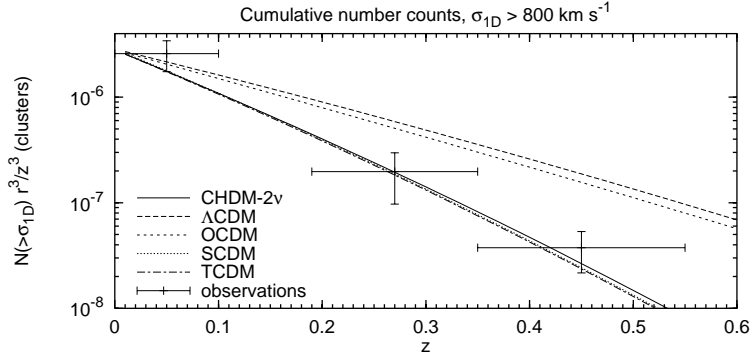}
  \caption{Abundance of clusters with $\sigma_{\rm 1D} > 800$ km s$^{-1}$.  All
           curves are Press-Schechter predictions normalized to agree with
           Fadda \etal\protect\cite{fadda:1996} at $z=0$.  The low-$z$ data
           point is from G97 and the other two are from CNOC.  We
           have removed assumptions of cosmology from the data by using
           velocity dispersions instead of masses, and by multiplying the
           number density by the cube of the ratio between the coordinate
           distance to $z$ and $cz/H_0$.}
  \label{fig:ndens}
\end{figure}
Figure~\ref{fig:ndens} shows the evolution in number density, where
all assumptions of cosmology on the observations have been removed.
The results favor a high-$\Omega_0$ cosmology, but there are many caveats
that weaken that conclusion.  These are discussed below.

\section{Caveats}
In arriving at our normalization and evolution results, we had to make
many assumptions.  We assumed throughout all the analysis presented here
that clusters were virialized objects.  However, it is well established that
clusters have remarkable substructures,\cite{west:1994} which represent the
signature of a lack of virialization.  We assumed clusters were spherical, but
highly elongated clusters are observed both in simulations and in Abell
clusters.\cite{binggeli:1982}  We made a very specific assumption about
the mass profiles of clusters --- that they were given by the
NFW profile.  Though this is a reasonable average case at
$r\sim r_{\rm vir}$, simulations show quite a lot of scatter around that
distribution.

Small statistics are also a problem, as only eight CNOC clusters survive their
$L_{\rm x}$ cut.  CNOC's correction for $L_{\rm x}$ selection is also
uncertain due to the large observed scatter in the
$\sigma_{\rm 1D}$--$L_{\rm x}$ relation.\cite{edge:1991a}

The definition of CNOC clusters in terms of a physical radius, while the
\emph{comoving} virial radius of a cluster of a given mass is slowly varying
with redshift, is a potential redshift dependent bias because a high redshift
object will include more unbound material than low redshift object of the same
mass.

There are further uncertainties in the Press-Schechter calibration, due to
cosmic variance.  Our choice of normalization at $M=5.5\times 10^{14}$ $h^{-1}$
$M_\odot$ is also arbitrary.  Different masses will yield somewhat different
normalizations.

\section{Conclusions}
Comparing cluster observations to simulations requires a great deal of
massaging.  It is preferable to perform most of the operations on the
simulation data because the uncertainties are smaller.  However, simulations
are very expensive, which makes exhaustive searches in parameter space
impossible.  So, we must use semianalytic techniques such as the Press-Schechter
approximation to ``extrapolate'' the models.  With a modified Press-Schechter
algorithm, we have renormalized all our models, and found that $\sigma_8$
is larger than Eke, Cole, \& Frenk \cite{eke:1996} if $\beta=1$.  Our result
of $\sigma_8\approx 0.6$ for $\Omega_0=1$ is consistent with $\beta=1.15$,
which is close to the value found by the Santa Barbara Cluster simulations.
\cite{evrard:1997}

Most importantly, using a plausible but non-unique set of assumptions,
we have found a counterexample to strong recent statements that cluster
number density evolution requires low $\Omega_0$.  The data is not currently
good enough to distinguish between reasonable values of $\Omega_0$.

%%%%%%%%%%%%%%%%%%%%%%%%%%%%%%%%%%%%%%%%%%%%%%%%%%%%%%%%%%%%%%
% Bibliography                                               %
%%%%%%%%%%%%%%%%%%%%%%%%%%%%%%%%%%%%%%%%%%%%%%%%%%%%%%%%%%%%%%

\bibliographystyle{unsrtnotitle} % for BibTeX - sorted numerical labels by
                                 % order of first citation. 
\section*{References}
\footnotesize
\bibliography{mnemonic,thesis}

\end{document}